\documentclass[aps,prd,11pt,nofootinbib,showkeys]{revtex4-2}

\usepackage{amsmath,amssymb,amsthm,mathrsfs,bbm}
\usepackage{latexsym,amscd,amsbsy,amsfonts,dsfont}
\usepackage{graphicx}
\usepackage{appendix}
\usepackage[utf8]{inputenc}
\usepackage{subfigure}  
\usepackage{slashed}
\usepackage{cancel}
\usepackage{multirow}
\usepackage{soul}
\usepackage{physics}
\usepackage{comment}
\usepackage[normalem]{ulem}
\usepackage[dvipsnames]{xcolor}
\usepackage[
      colorlinks=true,
      linktocpage=true,
      breaklinks=true,
      linkcolor=Aquamarine,
      urlcolor=Purple,
      filecolor=black,
      citecolor=Purple,
      menucolor=black,
      pdfstartview=FitV,
      bookmarksopen=true
      ]{hyperref}

\definecolor{emerald}{rgb}{0.31, 0.78, 0.47}
\definecolor{lila}{rgb}{0.78, 0.63, 0.78}

\numberwithin{equation}{section}

\newtheorem{theorem}{Theorem}[section]

\theoremstyle{definition}
\newtheorem{definition}{Definition}[section]

\theoremstyle{remark}
\newtheorem*{remark}{Remark}

\makeindex
\begin{document}

\title{Physical and Theoretical Challenges to Integrable Singularities}
\author{Julio Arrechea}
\email{julio.arrechea@sissa.it}
\affiliation{SISSA, via Bonomea 265, 34136 Trieste, Italy}
\affiliation{INFN (Sez. Trieste), via Valerio 2, 34127 Trieste, Italy}
\affiliation{IFPU, via Beirut 2, 34014 Trieste, Italy}
\author{Stefano Liberati}
\email{liberati@sissa.it}
\affiliation{SISSA, via Bonomea 265, 34136 Trieste, Italy}
\affiliation{INFN (Sez. Trieste), via Valerio 2, 34127 Trieste, Italy}
\affiliation{IFPU, via Beirut 2, 34014 Trieste, Italy}
\author{Hooman Neshat}
\email{hneshat@sissa.it}
\affiliation{SISSA, via Bonomea 265, 34136 Trieste, Italy}
\affiliation{INFN (Sez. Trieste), via Valerio 2, 34127 Trieste, Italy}
\affiliation{IFPU, via Beirut 2, 34014 Trieste, Italy}
\author{Vania Vellucci}
\email{vellucci@qtc.sdu.dk}
\affiliation{Quantum Theory Center ($\hbar$QTC) \& D-IAS, IMADA at Southern Denmark Univ.,\\ Campusvej 55, 5230 Odense M, Denmark}

\begin{abstract}
     Black hole spacetimes that exhibit integrable singularities have gained considerable interest as alternatives to both regular and singular black holes. Unlike most known regular black hole solutions, these models evade the formation of an inner horizon, thereby circumventing the well-known instability issues associated with such structures. Moreover, it has been suggested that the finite tidal forces near integrable singularities, may allow for a traversable extension beyond them. In this work, we present a set of arguments --- both theoretical, concerning test-field perturbations and the accumulation of matter at the singularity, and practical, related to the behavior of physical probes and extended objects --- with the aim of assessing the validity of the proposed integrability condition, and the feasibility of traversing such singularities. Our analysis highlights key subtleties that challenge the viability of said extensions as alternatives to regular black holes, and underscores the need for a more rigorous investigation of their physical implications.
\end{abstract}




\maketitle

\clearpage

\tableofcontents

\section{Introduction}
\label{Sec:Introduction}
Despite their widespread presence in popular culture, black holes remain a subject of intense debate among specialists. In their most idealized realizations, they are typically modeled as vacuum stationary solutions of general relativity. However, these solutions rely on mathematical idealizations, such as event horizons and singularities. Consequently, while they are often regarded as useful approximations to the dark and compact objects that we observe in nature, it is generally expected that the latter are ultimately characterized by local and regular properties.

One of the most striking issues with the stationary solutions of general relativity, is the teleological nature of black hole event horizons, i.e. the fact that one has to know the whole future of spacetime to specify them. As such, these objects are inherently undetectable in any finite-time experiment~\cite{Visser:2014zqa} and hence are useful mathematical abstractions rather than physical entities.

Similarly, spacetime singularities are widely interpreted as a signal to our incomplete understanding of the behavior of matter and gravity under extreme conditions~\cite{Crowther:2021qij}. At a fundamental level, both concepts hinge on the notion of infinity: event horizons require an infinite amount of time to be observed, while singularities correspond to infinite curvature, energy density, and matter concentration. It is generally expected that these infinities will be resolved by physics beyond general relativity, leading to a more physically meaningful description.

As a result, black holes in general relativity are best understood as mathematical constructs rather than complete physical models. While this framework has been remarkably successful, both theoretically and observationally, the search for a more realistic, non-singular description remains an active and well-motivated endeavor. In paraphrasing Stephen Hawking, this ``suggests that black holes should be redefined as metastable bound states of the gravitational field''~\cite{Hawking:2014tga}. Although a satisfactory non-singular alternative to classical black holes has yet to emerge, recent investigations have yielded significant partial results and insights. The search for alternatives to GR black holes can be motivated from a theoretical angle through the search of fully consistent alternatives~\cite{Ashtekar:2023cod}, or from the observational side to produce models that allow, at the very least, to test the black hole paradigm~\cite{Carballo-Rubio:2018jzw, Cardoso:2019rvt}. It will be the eventual merging of both approaches that will bring us close to a deeper understanding of the true nature of black holes.

As a step towards the search for theoretically well-motivated models, and stimulated by the recent advancements in black hole observations~\cite{LIGOScientific:2016aoc,EventHorizonTelescope:2019dse}, there has been a growing interest in the exploration of black hole metrics not endowed with singularities: regular black holes~\cite{Bardeen:1968qtr,Dymnikova:1992ux,Hayward:2005gi,Ansoldi:2008jw}. While varying in their characteristics, all ``regular black holes" (RBHs) are supported by the same physical principle, i.e., the presence of matter that violates the strong energy condition~\cite{Zaslavskii:2010qz}, thus allowing to bypass singularity theorems~\cite{1970RSPSA.314..529H}.

As a side note, we remark the importance of the fact that the concept of ``non-singular" is a multifaceted one in general relativity~\cite{GEROCH1968526,ellis1977singular, Curiel2023}, resulting in various regularity criteria adopted in the extant literature. A widely accepted criterion is that regularity should ensure the smooth evolution of physical trajectories and guarantee that observables remain finite along these paths. These requirements are typically encapsulated in the notions of geodesic completeness and predictability, which demand that physical observables remain finite throughout the spacetime.

RBHs that satisfy these conditions generally require that the expansion of outgoing radial null geodesics, which becomes negative inside a trapping horizon, must turn positive again before reaching the central region. This mechanism prevents the formation of focusing points and caustics, thereby avoiding singularity theorems. In contrast, the expansion of ingoing radial null geodesics may remain negative throughout, leading to the standard class of regular black holes with an untrapped core. Alternatively, if the expansion of ingoing null geodesics also changes sign, the resulting structures are known as ``black bounces", i.e.~solutions  characterized by a minimum radius that separates a trapped region from an anti-trapped one. For a thorough discussion of the above scenarios, see, e.g.~\cite{Carballo-Rubio:2019nel, Carballo-Rubio:2019fnb}.

Due to the expansion behavior described above, RBHs are characterized by at least two future trapping horizons: an outer and an inner one. In stationary geometries—such as the Bardeen or Hayward regular black holes—the inner horizon is a Cauchy horizon and, as such, is subject to classical instabilities, notably mass inflation~\cite{Poisson:1989zz,Poisson:1990eh,Ori:1991zz}, (while these are known to suffer from quantum instabilities as well~\cite{McMaken:2023uue}, we will leave them aside in our discussion). Although Cauchy-horizon instabilities are often considered beneficial in enforcing Strong Cosmic Censorship~\cite{Dafermos:2003wr, Dafermos:2002ka}, in the case of regular black holes, they could be catastrophic, potentially leading to the formation of a singularity at the inner horizon~\cite{Poisson:1989zz, Poisson:1990eh, Ori:1991zz, Dafermos:2017dbw}. Even though this may seem like an artifact of the global nature of a Cauchy horizon --- irrelevant to evolving black holes with finite lifespans --- recent work has shown that a (finite) exponential growth characterizes slowly evolving inner horizons~\cite{Carballo-Rubio:2024dca}. So, this instability extends also to astrophysical black holes (if they are regular or even simply rotating) as they settle into equilibrium after stellar collapse.

This is the main reason why there has been growing interest in the literature in black holes that occupy an intermediate status between fully singular ones --- where the metric is not even $C^0$  at the singularity \cite{sbierski2018c_0} --- and regular ones, where the metric is at least $C^2$ throughout. Black holes --- which are $C^0$ at the singularity but not $C^2$ --- exhibit what are commonly referred to as ``integrable singularities". Similarly to RBHs, integrable singularities entail non-vacuum spacetimes in general relativity, albeit they do not require the matter content to violate energy conditions as in their regular black holes and black bounce counterparts.

In other words, despite the presence of a curvature singularity at their core, these solutions describe geometries which are everywhere continuous and not endowed with inner horizons, as no untrapping of light rays occurs anywhere inside the outer horizon. Hence, black hole spacetimes with integrable singularities can be seen as an improvement over GR vacuum solutions, as they do not exhibit expected instabilities associated to Cauchy horizons (which are present not only in stationary RBH but also in the more mundane Reissner--Nordstr\"om or Kerr solutions). Furthermore, integrable singularities  are characterized by finite tidal forces acting on radial geodesics, implying the possibility to traverse the singularity and extend the spacetime through it~\cite{Lukash:2011hd}. However, it is worth noting that in \cite{Lukash:2013ts}, the authors explicitly discuss certain limitations concerning the traversability of such solutions, and emphasize that these configurations may still represent physically viable solutions in nature, despite potential issues related to traversability.

In this article, we aim to provide a more comprehensive perspective on the advantages and disadvantages of considering this class of objects as an alternative to RBHs. In particular, we examine the nature of these integrable singularities and assess their actual traversability and viability in physically realistic settings. While, integrable singularities appear in various contexts in which GR is modified at short distances/high energies~\cite{Lukash:2011hd,Casadio:2024zek,Daas:2025lzr,Bonanno:2016dyv,Bonanno:2017zen, Estrada:2024moz}, in what follows we shall analyze their properties as eternal, spherically symmetric solutions to GR, hence without relying on the features of the specific theory that might give rise to them.

The structure of the manuscript is as follows: In Sec.~\ref{Sec:Preliminaries} we review the geometrical properties of spacetimes with integrable singularities. Sec.~\ref{Sec: Strength of the singularity} discusses the strength of the singularities encountered by radial and non-radial geodesics, while Sec.~\ref{Sec:focusing point} pinpoints issues with the extendibility of spacetimes beyond integrable singularities. Sec.~\ref{Sec:integrability} analyzes scalar test-field perturbations, showing that, while the growth of stress-energy associated to spherical perturbations is consistent with integrability, the backreaction from non-spherical perturbations has a stronger divergence that could jeopardize the integrability of the background spacetime. Sec.~\ref{Sec:PhysicalObservers}, is devoted to extended infalling objects, examining the obstacles they face in attempting to traverse the singularity. We conclude with some further comments in Sec.~\ref{Sec: Conclusions}. Throughout this article, we shall adopt units where $c=G=1$, and the metric signature is $(-,+,+,+)$.

\section{Preliminaries}
\label{Sec:Preliminaries}

As anticipated, we restrict our discussion to eternal, spherically symmetric spacetimes, for which the metric reads
\begin{equation}\label{eq: static spherically symmetric}
    {\rm d}s^2=g_{tt}{\rm d}t^2+g_{rr}{\rm d}r^2+r^2({\rm d}\theta^2+\sin^2\theta {\rm d}\phi^2)\,,
\end{equation}
where $g_{tt}=-e^{\phi}g^{rr}$, with $\phi=\phi(r)$ a radial function, which we set to be zero. This choice restricts us to a subset of all possible spherically symmetric spacetimes. Despite so, this family of metrics includes Schwarzschild, Reissner-Nordstr\"{o}m, most regular black hole metrics, and black holes with integrable singularities. For simplicity, we assume $g_{tt}=-g^{rr}$ \footnote{One can trace this feature to the Ricci tensor property of having a vanishing radial null-null component, i.e. of being proportional to the metric in the t-r subspace~\cite{Jacobson:2007tj}.} in what follows, since this choice allows to analyze the main properties of integrable singularities in a more direct way. The metric acquires a more clear physical interpretation when written in terms of
\begin{equation}\label{eq: -g^{rr}}
    g^{rr}=1-\frac{2m(r)}{r}\,,
\end{equation}
where $m(r)$ is the so-called Misner-Sharp-Hernandez (MSH) mass function~\cite{Misner:1964je, hernandez1966observer} given by
\begin{equation}\label{MSH mass}
    m(r)=4\pi \int_{0}^{r}\epsilon(x)x^2 {\rm d}x\,,
\end{equation}
where $\epsilon(x)$ represents the proper energy density measured in the rest frame of the effective matter source. 

In spherical symmetry, the necessary condition for the absence of curvature singularities at $r=0$ is given by the finitude of the Kretschmann invariant
\begin{align}\label{eq: Kretchmann Scalar}
   \mathcal{K}
   &
   =R_{\alpha\beta\mu\nu}R^{\alpha\beta\mu\nu}=\frac{48m^{2}}{r^{6}}-\frac{64m'm}{r^{5}}+\frac{32\left(m'\right)^{2}}{r^{4}}+\frac{16m''m}{r^{4}}-\frac{16m''m'}{r^{3}}+\frac{4(m'')^2}{r^{2}}\,,
\end{align}
where the prime denotes differentiation with respect to $r$. By quick inspection of~\eqref{eq: Kretchmann Scalar} we see the Schwarzschild metric, with $m(r)=M$, has \mbox{$\mathcal{K_{\rm Schw}}=48M^{2}/r^{6}$}.

An energy density profile $\epsilon(x)$ compatible with regularity at $r=0$ requires that the MSH mass vanishes at $r=0$ as
\begin{equation}
    m(r)=m_{3}r^{3}+\order{r^{4}}\,,
\end{equation}
or faster, which is equivalent to $\epsilon(x)$ not blowing up as we approach $r=0$. In situations where the metric is selected to resemble a black hole, this scenario is well known to produce an inner horizon, as $g_{tt}$ changes sign twice when approaching the singularity from spatial infinity.
A particular example is given by Bardeen's spacetime
\begin{equation}\label{Eq:Bardeen}
    m_{\rm B}(r)=\frac{Mr^{3}}{\left(r^{2}+\ell^{2}\right)^{3/2}}\,, \quad \text{with}\quad  \mathcal{K_{\rm B}}=\frac{96 M^{2}}{\ell^{6}}+\order{r^{2}}\,.
\end{equation}
where $\ell$ is a regularization parameter that determines the scale of the quantum gravitational corrections responsible for the core.

However, as argued in~\cite{Casadio:2023iqt}, quantum effects might generate matter cores with divergent energy densities at $r=0$, resulting in mass functions of the form
\begin{equation}\label{eq:integrabilitycondition}
    m_{\rm Int}(r)=m_{1}r+\order{r^{2}}\,.
\end{equation}

\begin{figure}[htbp!]
    \centering
    \includegraphics[width=0.7\linewidth]{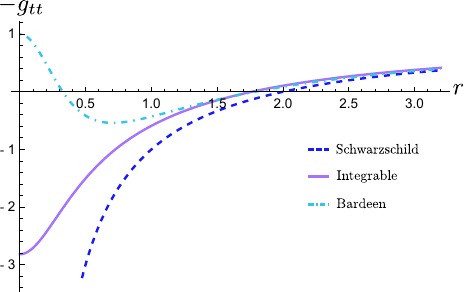}
    \caption{The behavior of $-g_{tt}$ is plotted for three different spacetime models. In dashed blue we plotted Schwarzschild spacetime, which exhibits a single horizon where $g_{tt}$ changes sign once. As $r\to 0$, $g_{tt}\to-\infty$. Bardeen spacetime~\eqref{Eq:Bardeen} with $\ell=1/2$ is shown in dot-dashed cyan and represents a regular black hole, where $g_{tt}$ changes sign twice, leading to an inner and outer horizon with a regular center. The integrable singularity model~\eqref{Eq:Integrable} with $C_{0}=3$, in solid purple, behaves as a separatrix case between the other two. It has a single horizon (like Schwarzschild) but the metric remains finite at $r=0$ (like Bardeen), although negative. In all plots we have chosen $M=1$.}
    \label{fig:gtt}
\end{figure}

If we no longer require $\epsilon(x)$ to be regular, but rather just ``integrable'', as to only give rise to a finite MSH mass, not only does the inner horizon disappear, but the metric components also remain finite at $r=0$. Taking, for example, the  metric proposed in~\cite{Casadio:2023iqt}, we have 
\begin{equation}\label{Eq:Integrable}
    m_{\rm Int}(r)=\frac{2M}{\pi}\arctan(\frac{C_{0}r}{M})\,, \quad \text{with}\quad \mathcal{K_{\rm Int}}=\frac{64C_{0}^{2}}{\pi^{2}r^{4}}+\order{r^{-2}}\,.
\end{equation}
where $C_{0}$ is a dimensionless constant related to the location of the horizon.
While the Kretschmann invariant is diverging, indicating the presence of a curvature singularity at $r=0$, at least its square root remains integrable which, as discussed in \cite{Casadio:2024zek}, is in agreement with the fact that $\epsilon \sim |\psi|^{2}\sim r^{-2}$ should be normalisable in the quantum theory ($\psi$ being the wavefunction of the static matter source). The distinction between the Schwarzschild, Bardeen~\eqref{Eq:Bardeen}, and the integrable black hole metric~\eqref{Eq:Integrable} is illustrated in Fig.~\ref{fig:gtt}.

Although this may already show an improvement over the Schwarzschild singularity, a more rigorous examination is required to determine whether the metric can truly be extended past $r=0$ and whether extended objects can traverse it without destruction. In the next section, we will introduce a formal definition of ``gravitational strength" for extended objects and develop the necessary framework to characterize the strength of an integrable singularity. These tools will be crucial for addressing the above-mentioned questions.

\section{Strength of integrable singularities}
\label{Sec: Strength of the singularity}

The concept of gravitational strength plays a crucial role in distinguishing between singularities that irreversibly destroy physical objects and those that may allow continued evolution beyond them. This distinction was first formalized by Ellis and Schmidt \cite{ellis1977singular}, and later refined by Tipler \cite{tipler1977singularities}, through the analysis of  an infalling geodesics' Jacobi fields  --- vector fields describing the infinitesimal separation between nearby geodesics and satisfying the geodesic deviation equation ---  when approaching a singularity. 

According to their definitions, a singularity is classified as strong if any extended object approaching it experiences a shrinkage to zero of the volume $V_{\rm dev}$, constructed from the aforementioned Jacobi fields, as this would inevitably lead to a complete destruction of the object. Conversely, a singularity is weak if objects can, in principle, survive passage through it. This is the so called Tipler's criterion in a nutshell.

Although Tipler's criterion provides a useful foundation to assess the strength of a singularity, subsequent studies pointed out some limitations \cite{nolan1999strengths, ori2000strength}. In particular, this criterion does not account for cases where the geodesic deviation stretches infinitely in one direction while collapsing in another, creating extreme deformations without necessarily crushing the entire ``volume element" $V_{\rm dev}$ to zero. 

In order to address this, Ori~\cite{ori2000strength} introduced a broader classification known as ``deformational strength", which encompasses both Tipler-strong singularities and those where {\em any} component of the Jacobi field diverges. This refined criterion has been instrumental in characterizing singularities within a range of physically relevant solutions. Let us then illustrate how $V_{\rm dev}$ is obtained.

Consider a timelike geodesic $\gamma(\tau)$ in a spacetime with metric $g_{\mu\nu}$. 
We parameterize $\gamma$ by the proper time $\tau$, and let $u^\mu(\tau)$ 
denote the 4-velocity (tangent vector) of the geodesic, so that
\begin{equation}
u^\mu = \frac{d x^\mu}{d\tau}\,, 
\qquad
g_{\mu\nu} \, u^\mu u^\nu = -1\,.
\end{equation}

A \emph{Jacobi field} along $\gamma(\tau)$ is a vector field $X^\mu(\tau)$ 
that satisfies the geodesic deviation equation
\begin{equation}\label{eq: geodesic deviation equation}
    u^{\nu}\nabla_{\nu}(u^{\alpha}\nabla_{\alpha}X^{\mu})=-R_{\nu \beta \alpha}^{\mu}u^{\nu}u^{\beta}X^{\alpha}\,,
\end{equation}
and represents an infinitesimal displacement
to a nearby geodesic. In $3+1$ dimensions, we can choose three linearly independent Jacobi fields 
$X^\mu(\tau),\, Y^\mu(\tau),\, Z^\mu(\tau)$ that are each orthogonal to $u^\mu(\tau)$, i.e.,
\begin{equation}
g_{\mu\nu} \, X^\mu u^\nu \;=\; 0, 
\quad
g_{\mu\nu} \, Y^\mu u^\nu \;=\; 0, 
\quad
g_{\mu\nu} \, Z^\mu u^\nu \;=\; 0.
\end{equation}
These three fields then span a small ``parallelepiped'' in the rest-space of the observer moving along $\gamma(\tau)$. Now, given a coordinate basis for the space orthogonal to the timelike geodesic $y^{a}\  (a=1,2,3)$, we can define a set of three spacelike tangent vectors to such space as
\begin{equation}
e^{\mu}_{a}=\left(\frac{\partial x^{\mu}}{\partial y^{a}}\right)_{\tau}
\end{equation}
where each $ e^{\mu}_{a}$ is spacelike and satisfies
\begin{equation}
g_{\mu\nu}\,e^\mu_{a}\,u^\nu=0,
\quad
g_{\mu\nu}\,e^\mu_{a}\,e^\nu_{b}=\delta_{ab}.
\end{equation}
The \emph{deviation volume} $V_{\rm dev}(\tau)$ is then defined via the orthonormal tetrad  {$\left\{u^\mu(\tau), e^\mu_{1}(\tau), e^\mu_{2}(\tau), e^\mu_{3}(\tau)\right\}$}. In particular, expressing each Jacobi field as
\begin{equation}
X^\mu(\tau) \;=\; X^1(\tau)\,e^\mu_{1}(\tau) \;+\; X^2(\tau)\,e^\mu_{2}(\tau) \;+\; X^3(\tau)\,e^\mu_{3}(\tau),
\end{equation}
and similarly for $Y^\mu(\tau)$ and $Z^\mu(\tau)$, 
then the \emph{deviation volume} $V_{dev}(\tau)$ is given by the determinant of the 
$3\times 3$ matrix of components
\[
V_{\rm dev}(\tau) 
\;=\; 
\det \begin{pmatrix}
X^1(\tau) & Y^1(\tau) & Z^1(\tau) \\[6pt]
X^2(\tau) & Y^2(\tau) & Z^2(\tau) \\[6pt]
X^3(\tau) & Y^3(\tau) & Z^3(\tau)
\end{pmatrix}.
\]
Alternatively, one can use the wedge product of the three Jacobi fields using the contracted Levi-Civita tensor $\epsilon_{\mu\nu\gamma}=\epsilon_{\mu\nu\gamma\delta}u^\delta$. 

Let us now recall Tipler's criterion that states if $V_{\rm dev}(\tau)$ goes to zero in a finite proper time, then the tidal forces (encoded in the curvature) are \emph{crushing} that volume element to zero. Ori's refinement of Tipler's criterion for a weak singularity additionally requires that none of the Jacobi fields' norms blow up in this limit, as this could result in a finite deviation volume in spite of an infinite stretching, and hence disruption, of the impinging object. We shall then say that the singularity is potentially ``traversable", if it is deformationally weak. Otherwise, matter cannot cross it.

It is worth emphasizing that the above defined volume element, $V_{\rm dev}$, is distinct from --- albeit related to --- the conventional cross-sectional volume element, $V_{\rm cross}=\sqrt{h}d^3y$ where $h_{ab}$ is the induced metric on the orthogonal space. Indeed, to deduce such relation it is sufficient to express the Jacobi fields 1-forms in terms of the tangent vectors (for a fixed $\tau \ :\ d\tau=0$) and the orthogonal space basis
\begin{equation}\label{eq: l in terms of e}
    \boldsymbol{X}_{i}=(X_{i})_{\mu}e^{\mu}_{a}dy^{a}\,.
\end{equation}
Once expressed in this way it is easy to see that
\begin{equation}\label{eq: }
    \begin{split}
        V_{\rm dev}&=
        \epsilon^{\mu\nu\gamma}X_{1 \mu}X_{2 \nu}X_{3 \gamma} \underbrace{\sqrt{h}\ dy^{1}\wedge dy^{2}\wedge dy^{3}}\\
        &=
         \epsilon^{\mu\nu\gamma}X_{1 \mu}X_{2 \nu}X_{3 \gamma}\  \underbrace{V_{\rm cross}}\,,
    \end{split}
\end{equation}
where $h$ is the determinant of the induced metric on the cross section $h_{ab}=g_{\mu \nu}e^{\mu}_{a}e^{\nu}_{b}$.

Appreciating this difference is crucial because, as we will see later, the calculation of the expansion parameter will force $V_{\rm cross}$ to go to zero at an integrable singularity, a result that holds for all central singularities. However, the more relevant volume element is $V_{\rm dev}$, which tells you the strength of the singularity and whether an object can ``traverse" the singularity or not. The latter can still satisfy the Tipler--Ori criteria, if some of the components of the Jacobi vector fields diverge at the singularity, so to keep $V_{\rm dev}$ finite, while their norms remain finite.

In \cite{Nolan:2000rn}, it was proven that for spherically symmetric metrics, all central singularities are deformationally strong for non-radial geodesics. However, in the case of radial geodesics, certain conditions can be checked to determine whether the singularity is deformationally weak or not. For an integrable singularity it is shown in \cite{Lukash:2011hd} that the tidal forces remain finite for freely falling particles, and it is checked in \cite{Estrada:2023dcj} that the singularity is deformationally weak for radial geodesics. These properties have led to the conclusion that the singularity can, at the very least, be reached by radial geodesics at $r=0$, and possibly even crossed without obstruction. Before addressing this statement carefully in Sec.~\ref{Sec:integrability} and Sec.~\ref{Sec:PhysicalObservers}, first we examine the extendibility of the metric in the next section.

\section{Integrable singularities are focusing points}
\label{Sec:focusing point}

\subsection{Expansions for null geodesics}\label{Subsec: expansions}
Let us take a closer look at the null geodesics that reach $r=0$. In particular, we compute the expansion parameter $(\Theta)$ for these geodesics, along with their corresponding affine parameter. While the expansion parameters are scalars, hence independent of the coordinate system in which they are expressed, they do depend on the choice of affinely parametrized vectors tangent to the geodesic congruence used to construct them. To obtain expansion parameters that are continuous throughout the horizon, we need to use coordinates that can be continuously extended across it. For this purpose, we write the metric~\eqref{eq: static spherically symmetric} in regular, Kruskal coordinates and, as usual, we start by considering the Eddington--Finkelstein null coordinates associated to light rays. Ingoing rays move along $v=$constant, and outgoing rays along $u=$constant curves, given by 
\begin{equation}\label{eq: u, v and tortoise}
    \begin{split}
    &u=t-r^{*},\\
    &v=t+r^{*},\\
    &r^{*}=\int \frac{{\rm d}r}{1-\frac{2m(r)}{r}}.
    \end{split}
\end{equation}
Since spherically symmetric spacetimes with integrable singularities present only one horizon $r_h$ at which $r^*$ diverges, we can define the Kruskal coordinates in the exact same way as we do in Schwarzschild
\begin{equation}
    U=-e^{-{u}/{4M}}, \quad{ }
    V=e^{{v}/{4M}}
\end{equation}
In these coordinates the metric reads
\begin{equation}
    ds^2=\alpha(r)\, {\rm d}U {\rm d}V e^{-{r}/{2M}}+r^2 {\rm d}\Omega^2
\end{equation}
where $\alpha(r)$ is a function that depends only on the choice of $m(r)$. These coordinates are regular at the horizon and can be extended through it: the initial region for $r>r_h$ is described by $U<0$ and $V>0$, extending the range of these coordinates to $U>0$ and $V<0$ allows to cover also the interior region $r<r_h$. 

The affinely parametrized tangent vector to outgoing null geodesics ($U=$constant) is:
\begin{equation}
    k_{\mu}=-\partial_{\mu} U\,,
\end{equation}
whose expansion is
\begin{equation}
    \Theta_{+}=\nabla^{\mu}k_{\mu}= -\frac{U}{2Mr}\,,
\end{equation}
which is, as expected, positive outside the horizon, negative inside the trapped region ($r<r_h$, $U>0$) and diverges at $r=0$. The same calculation for ingoing null geodesics also shows a negative divergence at $r=0$. This means that $r=0$ is a \emph{focusing point}.

As usual, $\Theta$ can be expressed as the fractional rate of change of $V_{\rm cross}$~\cite{Padmanabhan:2010zzb}
\begin{equation}\label{eq:thetatoV}
    \Theta=\frac{1}{ V_{\text{cross}}}\frac{d}{d \tau} V_{\text{cross}}\,,
\end{equation}
and so, given that $\Theta\to-\infty$ as we approach the singularity, the cross-sectional volume (for both ingoing and outgoing congruences) goes to zero $V_{\text{cross}}\to0$, as we anticipated at the end of the previous section.


An essential request for declaring such focusing point a singularity, is that null geodesics reach this focusing point in a finite value of the affine parameter. In order to show this, we shall follow the approach outlined in \cite{visser2023efficient}. 

In a static, spherically symmetric spacetime, the metric can always be expressed in a block-diagonal form by choosing appropriate coordinates
\begin{equation}\label{eq:blockmatrix}
g_{\mu\nu}=\left(
\begin{array}{c|c}
-N^2 & 0 \\ \hline
0 & g_{ij}
\end{array}
\right)\,,
\end{equation}
where all metric components remain time-independent. For any null geodesic, we impose the null condition on the line element
\begin{equation}\label{eq:null curve}
    -N^2 {\rm d}t^2 + g_{ij}{\rm d}x^{i}{\rm d}x^{j}=0 \,,
\end{equation}
which means (up to an irrelevant sign)
\begin{equation}\label{eq: N dt}
    N {\rm d}t=\sqrt{g_{ij}{\rm d}x^{i}{\rm d}x^{j}}\,.
\end{equation}
On the other hand, the Killing vector associated with time translations in a static spacetime is given by
\begin{equation}\label{eq: Killing static}
    \xi_{\mu}=(-N^2,0,0,0)\,.
\end{equation}
Contracting this with the four-velocity of a null geodesic, we obtain a conserved quantity, specifically, the energy per unit affine parameter:
\begin{equation}\label{eq: constant of motion null}
    -N^2 \frac{{\rm d}t}{{\rm d}\lambda}=\mbox{constant}=E\,,
\end{equation}
which can be used to find an expression for $d\lambda$, where $\lambda$ is the affine parameter. 
Therefore, up to some undetermined non-zero constant, we have
\begin{equation}\label{eq:dlambda}
    {\rm d}\lambda \propto N\sqrt{g_{ij}{\rm d}x^{i}{\rm d}x^{j}}\,.
\end{equation}

Using the static and spherically symmetric metric introduced before~\eqref{eq: static spherically symmetric}, we substitute the metric into the expression for $d\lambda$, yielding
\begin{equation}\label{eq: affine parameter integrable}
    {\rm d}\lambda\propto\sqrt{{\rm d}r^2 + (r^2 -2m(r)r){\rm d}\Omega^2}\,.
\end{equation}
for radial geodesics $d\Omega=0$, so this simplifies to 
\begin{equation}\label{eq:lambda}
    \lambda\propto r\,.
\end{equation}
Thus, for a radial null geodesic starting from some finite radius and falling toward the singularity, it takes a finite amount of affine parameter to reach $r=0$ which, as proven above, is also a focusing point.


\subsection{Conjugate points and extendibility}\label{Subsec: Conjugate}

In order to study geodesics that end in the singularity and better understand the issues in their potential extension, we can rely on the notion of \textit{conjugate points}. As discussed in detail in \cite{wald2010general}, conjugate points are formally defined as follows:
\begin{definition}
    Consider a pair of points $p$ and $q$ on a geodesic $\gamma$. For this pair, if there exists a Jacobi field $X^{\mu}$ which is not identically zero but vanishes at both $p$ and $q$, then this pair is called \textit{conjugate}. 
\end{definition}

\begin{remark}\label{remark:conjugate}
    We consider a timelike/null\footnote{There are subtle differences between timelike and null geodesics which are not relevant to our current argument. For a more detailed discussion, see \cite{wald2010general}.} geodesic $\gamma$ and point $p$ belonging on this geodesic. We take the congruence of all timelike/null geodesics passing through $p$, so that every Jacobi field which vanishes at $p$ is a deviation vector for this congruence. It is shown in \cite{wald2010general} that a point $q$ which lies on $\gamma$ in the future of $p$, is conjugate to $p$ if and only if the expansion $\Theta$ of such a geodesic congruence approaches $-\infty$ at $q$. 
\end{remark}

At this point, there is an important clarification to make. That $r=0$ is a focusing point for null geodesics does not imply that it is a conjugate point to a given point $p$ preceding it. The reason is that conjugate points must belong to the manifold, and in most cases, singularity theorems --- such as the original theorem by Penrose \cite{penrose1965gravitational} --- establish that such a point cannot be part of the manifold. Consequently, under the assumptions of the Penrose theorem, $r=0$ cannot technically be considered a conjugate point because it is a singularity. The existence of conjugate points is a property of complete geodesics, whereas geodesics terminating at $r=0$ are incomplete.

This suggests that in order to prevent $r=0$ from being excluded from the manifold, at least one of the assumptions of Penrose's singularity theorem must be violated without introducing any additional restrictive assumptions. In order to see which one, let us assume that $r=0$ is included as a point in the extended spacetime, then according to the Remark~\ref{remark:conjugate}, $r=0$ would be a conjugate point to any other point inside the trapped region. This follows from the fact that any two infinitesimally close geodesics passing through a point inside the trapped region will inevitably intersect again at $r=0$, where the expansion parameter blows up negatively.

By itself, the existence of conjugate points does not indicate any fundamental issues with the spacetime or the geodesics that contain them. Rather, conjugate points simply signify the failure of timelike geodesics to maximize proper time between two points and also imply that null geodesics can be smoothly deformed into timelike ones if a conjugate point exists between the two endpoints \cite{wald2010general}. Thus, while conjugate points are an important concept in the study of geodesic behavior, their presence alone does not imply any pathological feature of the spacetime.

However, we have just seen that $r=0$ of a traversable singularity would be a special point. It would not be merely a conjugate point for some specific geodesics or a particular family of geodesics; rather, all curves inside the trapped region inevitably focus at $r=0$. In this sense, $r=0$ functions as a ``universal'' conjugate point. This very fact nonetheless, would clash with the global hyperbolicity of spacetime. This can be seen using the following theorem~\cite{wald2010general} 
\begin{theorem}
    Let $(M,g_{\mu \nu})$ be a globally hyperbolic spacetime and let $K$ be a compact, orientable, two-dimensional spacelike submanifold of $M$. Then every $P\in \dot{I}^{+}(K)$ lies on a future directed null geodesic starting from $K$ which is orthogonal to $K$ and has no point conjugate to $K$ between $K$ and $P$.
\end{theorem}

This implies that if the spacetime were to be extended beyond $r=0$, global hyperbolicity--- one of the assumptions of Penrose's theorem --- would necessarily be lost. The reason is that any causal curve originating from a two-dimensional spacelike surface inside the trapped region would have to pass through $r=0$ to reach the extended region. However, at the same time, $r=0$ has been shown to be a conjugate point to every point inside the trapped region. This suggests that $r=0$ behaves as a ``Cauchy point'' \footnote{Throughout this manuscript, when referring to $r=0$ as a ``point," we mean a point in the spatial geometry (i.e., on a spatial slice), not a point in the full spacetime. Indeed, once $r=0$ is included in the spacetime extension, it can be consistently regarded as a point on each constant-time hypersurface.}, which could be interpreted as a Cauchy horizon shrunk into a single point.

This violation of global hyperbolicity may initially appear problematic in the context of the Strong Cosmic Censorship Conjecture. In its original format, the conjecture states that \textit{maximal future Cauchy developments to generic asymptotically flat initial data are future inextendible as a suitably regular Lorentzian manifold} \cite{Penrose:1980ge}. A crucial subtlety is in what sense we disallow extensions --- i.e., how “smooth” or “regular” must a putative extension be in order to violate the conjecture? Over the decades, several versions have been proposed. Most notably, there are the “modern” $C^{0}$ and $C^{2}$ formulations of the conjecture, as well as other variants which exist in the literature\footnote{For a more comprehensive discussion on the Strong Cosmic Censorship Conjecture, check \cite{van2025strong, landsman2021singularities, Dafermos:2017dbw, Cardoso:2017soq}}.

However, as we have seen before, at $r=0$, curvature invariants such as the Kretschmann scalar diverge, ensuring that the metric is $C^{2}$-inextendible \cite{o1983semi}. This fact saves the day, as it makes the solution compatible at least with the $C^{2}$ formulation of the SCC. In comparison, the situation is somewhat similar in the case of regular black hole solutions containing a Cauchy horizon. Mass inflation instabilities \cite{Poisson:1989zz, Poisson:1990eh, Ori:1991zz} render the Cauchy horizon $C^{2}$ inextendible \cite{Dafermos:2003wr, Dafermos:2002ka}, therefore ensuring the survival of the $C^{2}$ formulation of the Strong Cosmic Censorship Conjecture.

While the mechanism enforcing $C^2$-inextendibility is different (mass inflation at a Cauchy horizon versus curvature divergence at $r=0$), the outcome is analogous. The metric remains continuous, and finite tidal forces suggest that a $C^{0}$ or piecewise $C^{1}$ extension could be possible\footnote{For more discussions regarding low regularity extensions, check \cite{hawking2023large, clarke1993analysis}}. Thus, from the standpoint of $C^2$-inextendibility, regular black hole models with Cauchy horizons and black hole models with integrable singularities share a common feature: neither allows a smooth extension, but both might admit a weaker one. In this sense, they stand on an equal footing in their agreement with the $C^2$ formulation of the conjecture.

\section{Threats to the integrability of the solution}
\label{Sec:integrability}

So far, it appears that an integrable singularity, acting as a ``Cauchy point", shares certain similarities with a Cauchy horizon. However, as noted earlier, Cauchy horizons are known to suffer from instabilities. This raises an important question: If an integrable singularity proves to be stable and its integrability conditions remain preserved over time, could it serve as a viable alternative to spacetimes containing Cauchy horizons?

In \cite{Casadio:2024zek}, it is argued that in the presence of an integrable singularity, a ``quantum core" can replace the traditional singularity in both static and rotating solutions. In both cases, the source mass near  $r=0$ is expected to exhibit a linear dependence on $r$:
\begin{equation}\label{eq: m(r)}
     \lim_{r\to0} m(r)=m_{1} r+O(r^2)\,. 
\end{equation}
Recall that the MSH mass function is proportional to the energy density present in the spacetime~\eqref{MSH mass}. This condition is essential to maintain the continuity of the metric components, as well as to ensure the integrability of certain physical quantities. Any term with a power of $r$ smaller than unity would spoil the continuity of the metric at $r=0$.  

In Section.~\ref{Sec: Strength of the singularity}, we discussed how under certain conditions radial geodesics may admit an extension beyond the singularity, whereas non-radial geodesics are inevitably prevented from crossing $r=0$. This implies that any matter falling toward the singularity with non-zero angular momentum will accumulate at $r=0$, contributing to the mass of the quantum core. This accumulation occurs due to a key factor: $r=0$ is a focusing point, causing all geodesics to converge there. Consequently, any particle that cannot traverse $r=0$, also lacks the option of reflecting back, leading to an inevitable accumulation of matter at the singularity.

Since there is no restriction on the density distribution of infalling matter, the mass of the quantum core can change in a non-trivial way
\begin{equation}\label{eq: change in core's mass}
    \delta m= m(r) + \delta\Bar{m}(r,\theta,\phi)\,.
\end{equation}

In principle, this need not pose a problem if we considered a less symmetric spacetime to begin with, as for example, a generalized $m(r,\theta,\phi)$. However, it raises suspicion onto whether integrability is just a feature of spacetimes enjoying a high degree of symmetry that will not be preserved (or be present only for a narrow set of geodesic observers) in more generic situations. Still, we could conjecture potential mechanisms that prevent this loss of symmetries. 

One possibility is the existence of an ``isotropization" process within the quantum core, which ensures that the mass function adjusts itself to maintain the strict conditions that integrability requires. For example, a mechanism could exist that dissipates or eliminates angular momentum before the infalling matter reaches $r=0$, enforcing a purely radial infall. Since it remains unclear how this non-trivial accumulation of matter could be prevented, the very nature of the singularity --- specifically, its strength for non-radial infalling matter --- poses a serious challenge to the stability of the integrability property of the core.

To address the robustness of the integrability condition in a more rigorous way, we shall then examine the behavior of scalar test field perturbations near the singularity, and calculate the energy-density associated to these perturbations.

\subsection{General test field perturbations}
\label{Subsec: perturbations}

The behavior of linear test field perturbations\footnote{For a reference to non-linear analysis, see e.g.~\cite{burko1997late}.} inside trapped regions, with particular emphasis on the Schwarzschild metric, has been studied in \cite{doroshkevich1978spacetime}. The analysis performed in that article revealed that, as we approach the point $r=0$, the solution for massless scalar fields displays a logarithmic blow-up $\propto \ln r$, and fields with nonzero spin grow in proportion to an inverse power of $r$. These divergent solutions point towards the need to go beyond test-field approximations and consider the backreaction produced by the stress-energy tensor associated to said fields, which might drastically affect the background spacetime.

The features of these test field perturbations are dependent on the shape of the spacetime near the singularity. Here, we follow similar arguments to analyze test fields near $r=0$ in the case of generic (spherically symmetric) spacetimes with integrable singularities. Recall that we are working in a static spherically symmetric background,
\begin{equation}\label{eq: static, spherically symmetric metric}
    {\rm d}s^{2}=-f(r){\rm d}t^2+\frac{1}{f(r)}{\rm d}r^{2} +r^2({\rm d}\theta^2 +\sin^2\theta \ {\rm d}\phi^2)\,,
\end{equation}
with $f(r)=1-2m(r)/r$. Massless fields can be decomposed in spherical harmonics and Fourier modes
\begin{equation}\label{eq: field decomposition}
    \Phi_{s}(t,r,\theta,\phi)=\sum_{l,m}\frac{1}{r}\Psi_{s,k l}(t,r)Y_{lm}(\theta,\phi)= \int dk  \sum_{l,m}\frac{1}{r}e^{- i k t}\Psi_{s,k l }(r)Y_{lm}(\theta,\phi)\,,
\end{equation}
here $s$ represents the spin, and the radial modes $\Psi_{s}(r)$ (we drop the $k,~l$ dependence for brevity) satisfy the wave equation
\begin{equation}\label{eq: wave equation for mode amplitudes}
    \frac{d^2 \Psi_{s}}{dr_{*}^{2}}+(k^2-V_{s})\Psi_{s}=0\,.
\end{equation}
Because we are considering massless fields, the standard frequency term $\omega^2$ in the wave equation is replaced by $k^2$, which is the wave number $(\omega^2=k^2)$. The tortoise coordinate is again defined as~\eqref{eq: u, v and tortoise}, and the potential depends on the spin-weight of the perturbation and the metric functions
\begin{equation}\label{eq: potential}
     V_{s} = f(r) \left[ \frac{l(l+1)}{r^2} + \frac{2(1 - s^2)m(r)}{r^3} - (1 - s)\left( \frac{2m'(r)}{r^2} \right) \right]\,.
\end{equation}
We write~\eqref{eq: wave equation for mode amplitudes} for $s=0$ fields as
\begin{equation}\label{eq: wave equation spin zero}
    \frac{d^2 \Psi_{0}}{dr_{*}^{2}}+\left\{k^2-\left(1-\frac{2m(r)}{r}\right) \left[ \frac{l(l+1)}{r^2} + \frac{2m(r)}{r^3} -\left( \frac{2m'(r)}{r^2} \right) \right]\right\}\Psi_{0}=0\,.
\end{equation}

In what follows, we will cast this equation in a form more suitable to find its solutions near $r=0$ via the Frobenius method, that is we write the derivatives in terms of $r$ using~\eqref{eq: u, v and tortoise}. From~\eqref{eq: wave equation spin zero}, it is straightforward to write
\begin{equation}
   \Psi_{0}''+\frac{\mathcal{P}}{r}\Psi_{0}'+\frac{\mathcal{Q}}{r^{2}}\Psi_{0}=0,
\end{equation}
where $\mathcal{P}$ and $\mathcal{Q}$ are the radial functions
\begin{align}\label{Eq:Frob}
    \mathcal{P}=\frac{r f'}{f}\,,\qquad \mathcal{Q}=\frac{r^{2}k^{2}}{1-2m/r}-\frac{l(l+1)}{1-2m/r}+\frac{2\left(m'-m/r\right)}{1-2m/r}\,.
\end{align}
We assume the metric displays an integrable singularity at $r=0$ and locally obeys the expansion
\begin{equation}
    m(r)=m_{1}r+m_{2}r^{2}+\order{r^{3}},
\end{equation}
hence we find
\begin{equation}
    \mathcal{P}=-\frac{2m_{2} r}{1-2m_{1}}+\order{r^{2}},\quad \mathcal{Q}=-\frac{l(l+1)}{1-2m_{1}}+\frac{2m_{2}r\left[l\left(l+1\right)-1+2m_{1}\right]}{\left(1-2m_{1}\right)^{2}}+\order{r^{2}},
\end{equation}
which implies that $r=0$ is a regular singular point for this second-order differential equation\footnote{For a second order differential equation of the form $y''+p(x)y'+q(x)y=0$, $x = x_0$ is called a \emph{regular singular point} if 
$(x - x_0)\,p(x)$ and $(x - x_0)^2\,q(x)$
are analytic (i.e., admit convergent power-series expansions) in a neighborhood of $x_0$. }, which guarantees that we can find the solution near $r=0$ using the Frobenius method.

The solutions to~\eqref{Eq:Frob} near $r=0$ can be written as
\begin{equation}\label{eq: Spin zero, solution}
    \Psi_{0}(0)=r^{R}\sum_{n=0}^{\infty}a_{n}r^n\,.
\end{equation}
with $n$ being integer powers, and $R$ being powers which are not necessarily integers\footnote{$R$ powers can be rational, irrational, or even complex numbers.}, which are solutions to the indicial equation 
\begin{equation}\label{eq: Spin zero, indicial}
    R(R-1)+\mathcal{P}(r=0)R+\mathcal{Q}(r=0)=0\,.
\end{equation}
We find two independent roots for $R$
\begin{equation}\label{eq: Spin zero, R}
    \begin{split}
        R_1=\frac{1}{2}\left(1-\frac{\sqrt{-4l(l+1)+\alpha}}{\sqrt{\alpha}} \right),\quad 
        R_2=\frac{1}{2}\left(1+\frac{\sqrt{-4l(l+1)+\alpha}}{\sqrt{\alpha}} \right)\,,
    \end{split}
\end{equation}
where we have called $1-2m_{1}=-\alpha$ (with $\alpha$ being a positive number, since $m_{1}>\frac{1}{2}$ as discussed in \cite{Casadio:2023iqt}). All our considerations below are restricted to the leading-order contributions as they are the ones that will dominate the backreaction.

\paragraph{s-waves:} For the case of $l=0$, we have $R_1=0$ and $R_2=1$. Since the roots differ by an integer number, we can write the two independent solutions as
\begin{equation}\label{eq: Spin zero, roots}
    \begin{split}
        &\Psi_{0}^{1}=r\sum_{n=0}^{\infty} a_{n}r^n\,,\\
        &\Psi_{0}^{2}= \ln(r)\ r \sum_{n=0}^{\infty} a_{n}r^n\ + \sum_{n=0}^{\infty}b_{n} r^{n}\,,\\
        &\Psi_{0}=A_{1}\Psi_{0}^{1}\ +\ A_{2} \Psi_{0}^{2}\,.
    \end{split}
\end{equation}
therefore, the resulting solution, keeping only the leading terms ($n=0$), would be
\begin{equation}\label{eq: Spin zero, solution leading term }
    \Psi_{0}\propto r+ r\ln(r)\,.
\end{equation}
This solution describes the asymptotic behavior of the mode amplitudes near $r=0$ and hence for $r\to 0$ and constant $t$ one has for the total field \eqref{eq: field decomposition}
\begin{equation}\label{eq: Spin zero field}
    \Phi_{0}(r)\propto \ln(r)\,.
\end{equation}
which is the same as what was found in \cite{doroshkevich1978spacetime}, in the case of Schwarzschild.

Before moving on to non s-wave solutions, a few comments are in order:

\begin{itemize}
\item It is easy to see that at leading order the solution \eqref{eq: Spin zero field} is independent of $k$. In order to recover the dependence on $k$, one should put the full solution~\eqref{eq: Spin zero, roots}, back into~\eqref{eq: wave equation for mode amplitudes}, and find the relevant recursion relations for $a_{n}$ and $b_{n}$, which would depend on $k$. The full physical field $\Phi_{0}$, can then be calculated, taking the integral over $k$ in~\eqref{eq: field decomposition}\footnote{A calculation for the case of a collapsing star has been done in \cite{patashinskil1974damping}.}. Lastly, the coefficients $A_{1}$ and $A_{2}$ are related to the corresponding initial conditions of the field at early times, i.e., at horizon-crossing\footnote{There remains the possibility that evolving a physical field from early times results in a vanishing of the $A_{1,2}$ coefficients in~\eqref{eq: Spin zero, roots} for all values of $l$ and $k$. We have not ruled out this possibility, but consider it quite unlikely unless initial conditions are fine tuned.}. One can choose initial conditions equivalent to the ones chosen in \cite{doroshkevich1978spacetime}.
\item The analysis in~\cite{doroshkevich1978spacetime} is interested in the behavior of fields in the trapped region at late times after the black hole formed. In this regime, the dominant contribution to the field arises from long-wavelength modes, corresponding to the limit $k\to 0$ in Fourier space. This is due to the strong redshifting of out-going modes as they propagate inside the black hole.  As a result, the relevant part of the solution can be captured by expanding around small $k$, which results in tails that decay polynomially in $t$, reproducing the interior counterpart of Price's law~\cite{Gundlach:1993tp}. In our case, the late-time behavior of fields is not our main concern but rather their blow-up as they approach $r=0$, independent of their behavior in $t$. 
\end{itemize}

\paragraph{Non s-waves:} For $l>0$ on the other hand, the scalar field written with only the leading terms ($n=0$) would be
\begin{equation}\label{eq: Spin zero, solution leading term}
    \Phi_{0}(r)\propto r^{R_{1}-1}+ r^{R_{2}-1}\,,
\end{equation}

An analysis of this class of solutions considering all the possible values for $l$ and $\alpha$ in~\eqref{eq: Spin zero, R} is shown in the Appendix~\ref{sec: appendix}. Regardless of the values of these parameters, there exists a leading term blowing up near $r=0$ with an inverse power law of $r$, namely
\begin{equation}\label{eq: spin zero, field power law}
    \Phi_{0}(r)\propto r^{-\beta}\,,\quad \mbox{with}\quad\beta\geq\frac{1}{2} \,.
\end{equation}

We can now calculate the energy density associated with the field and see how it behaves as a function of $r$. We start by writing the energy-momentum tensor for a massless scalar field
\begin{equation}\label{eq: Spin zero, energy momentum}
    T_{\mu\nu}=\partial_{\mu}\Phi_{0}\partial_{\nu}\Phi_{0}-\frac{1}{2}g_{\mu\nu}g^{\alpha\beta}\partial_{\alpha}\Phi_{0}\partial_{\beta}\Phi_{0}\,,
\end{equation}
then we contract it twice with the four-velocity of a radially infalling observer 
\begin{equation}\label{eq: four-velocity, radially infalling}
u^{\mu}=\left(\frac{1}{1-\frac{2m(r)}{r}},-\sqrt{\frac{2m(r)}{r}},0,0\right)\,,
\end{equation}
so to obtain the energy density measured by that observer near $r=0$
\begin{equation}\label{eq: Spin zero, energy-density}
        \rho=T_{\mu\nu}u^{\mu}u^{\nu}
            =T_{tt}u^{t}u^{t}+T_{rr}u^{r}u^{r}
            =\left(1+\frac{\alpha}{2}\right)\left(\partial_{r} \Phi_{0}\right)^2\,.
\end{equation}
this suggests that since for $l=0$ the field blows up as $\Phi_{0}\propto \ln(r)$, then
\begin{equation}\label{eq: Spin zero}
   \lim_{r\to 0} \rho\sim \frac{1}{r^2}\,,
\end{equation}
which is blowing up but is still integrable. 
However, for the case of $l>0$, the field blows up as $\Phi_{0}\propto r^{-\beta}$, therefore
\begin{equation}\label{eq: Spin zero}
    \lim_{r\to 0}\rho\sim r^{-2(\beta+1)}\,,
\end{equation}
and because $\beta\geq\frac{1}{2}$, the energy density is no longer integrable
\begin{equation}
    \int_{0}^{r}\rho\ r^{2}\sin{\theta}\, {\rm d}r\, {\rm d}\theta \,{\rm d}\phi =\infty\,.
\end{equation}

This analysis shows that the integrability condition, which is built into the solution for physical quantities associated with the source --- such as the energy density --- fails to hold for a general field perturbation considered on top of the background. In other words, although the background solution is constructed to ensure the integrability of certain quantities, this feature does not necessarily extend to perturbative fields evolving on the same geometry.
 
Lastly, it is instructive to contrast this behavior with what occurs in the cores of non–black hole solutions, such as stars, in which $r=0$ is a timelike surface. In those cases, the equivalent to~\eqref{Eq:Frob} also has a regular singular point and the indicial polynomial equation has the roots
\begin{equation}
    R^{\rm star}_{1}=l+1,\quad R^{\rm star}_{2}=-l.
\end{equation}
For $l>0$, there are divergent solutions but, since $r=0$ is not inside a trapped region, the field equation must be solved as a boundary-value problem and, in this case, the physically reasonable condition that must be imposed is that the field vanishes at $r=0$, hence we can discard the singular solution from the start and the test-field approximation remains valid.

In contrast, in the black hole interior, we are faced with an initial-value problem due to the fact that the roles of $t$ and $r$ swap. The mode equation is solved by imposing initial conditions at some surface in the far past (for example, upon horizon-crossing~\cite{doroshkevich1978spacetime}).
As a result, the freedom to impose reflective boundary conditions at $r=0$ --- commonly used to enforce regularity in stellar interiors --- is no longer available. In this context, $r=0$ is not a point in space but rather a moment in time that indicates when the focusing of all infalling matter into a single point takes place. Integrable singularities exhibit a discrepancy between the properties of the stress-energy source generating the background metric and the energy content associated to physical fields, which tends to diverge faster. This is a direct consequence of the presence of the focusing point and explains why integrable singularities might not survive in realistic situations.

We emphasize that the present study is fully classical: both the background geometry and the test-field perturbations are treated within classical general relativity. Although this framework is sufficient to reveal serious problems with integrability, it leaves out quantum effects that could become crucial near the core, where curvature scalars grow and a smooth classical energy-density description may no longer apply. Recent works such as \cite{casadio2023quantum}, highlight that a fully quantum treatment --- in which the core is defined probabilistically, akin to the charge cloud of a hydrogen atom --- may be required in order to decide whether the breakdown of integrability, that we have uncovered at the classical level persists, is alleviated, or is replaced by possible new quantum phenomena.

Finally, it is worth mentioning that an alternative interpretation of our analysis is that significant departures from spherical symmetry are likely to be generically induced by matter perturbations. While we cannot exclude the possibility that stable integrable singularity solutions might exist beyond spherical symmetry, our analysis suggests that realizing such solutions in full generality may be difficult, given that arbitrary perturbations could impinge upon the singularity. Nonetheless, we hope that our investigation will encourage further scrutiny of such scenarios.

\section{Integrable singularities for physical observers}
\label{Sec:PhysicalObservers}
We have shown that the gravitational effects of scalar fields will generically modify integrable singularities in ways that might destroy integrability. Even if, eventually, some backreaction mechanism was able to preserve integrability in the non-linear theory, these singularities would still pose problems to physical (extended) observers as they are approached. Before concluding this article, let us include here some comments and reflections on these aspects.
\subsection{Deviations from geodesic motion}
\label{Subsec: Deviations from geodesic motion}

In realistic physical scenarios, extended objects, unlike idealized point particles, experience internal stresses and forces that influence their motion in strong gravitational fields. While a point particle in general relativity follows a geodesic when no external forces act on it, an extended body obeys a more complex set of equations, such as the Mathisson--Papapetrou--Dixon (MPD) equations \cite{Mathisson:1937zz, papapetrou1951spinning, dixon1974dynamics}. These equations describe the evolution of a spinning test body in curved spacetime and account for the coupling between the body's internal angular momentum and the spacetime curvature. As a result, different points within the object generally do not follow the same geodesic trajectory \cite{Cotton:2021tfl}, leading to internal deformations, torque, and even shifts in the object's center of mass. Furthermore, in certain conditions --- such as in highly curved regions near a singularity --- even the center of mass itself may deviate from a geodesic path due to spin-curvature coupling and tidal effects.

This complexity becomes particularly relevant when considering the nature of singularities. For instance, in~\cite{Nolan:2000rn} all the discussions and classifications are for geodesic motion. For an extended object, this assumption is no longer valid. The MPD equations introduce non-trivial corrections to the motion, meaning that different parts of the object may follow non-radial trajectories even if the overall motion appears radial in a naive point-particle approximation. Consequently, internal forces may shift individual elements of the object onto geodesics that terminate in a strong singularity, undermining the argument that the singularity is weak.

This observation highlights a key limitation of singularity analyses based purely on geodesic motion. The existence of conditions under which a singularity is weak for radial geodesics but strong for non-radial ones is already a highly fine-tuned result. However, because real physical objects are extended and not constrained to geodesic motion, this fine-tuning becomes even more fragile. Even a small perturbation due to internal forces could shift parts of the object onto trajectories that encounter a strong singularity, leading to its complete destruction. Therefore, the conclusions drawn in the geodesic-based analysis may not extend to realistic astrophysical bodies, which are subject to a more complex and less predictable interaction with the singularity.

A key difference between the structure of an integrable singularity and that of a Cauchy horizon lies in the nature of their respective strengths. In \cite{burko1999strength}, it has been shown that the null singularity --- possibly produced by a mass inflation instability at the inner/Cauchy horizon --- remains weak for all geodesics. This universality makes the possibility of crossing a Cauchy horizon without being torn apart much more plausible. Since real physical objects have internal forces that may approximate different types of geodesic motion, a Cauchy horizon that is weak for all geodesics offers far more flexibility in allowing an extended object to pass through it intact. Even if the center of mass follows a non-radial trajectory due to spin-curvature coupling or other perturbations, the weakness of the singularity remains robust across all possible geodesic approximations of the motion.

\subsection{Energy encountered by a free-falling observer}
\label{Subsec: Energy encountered by a free-falling observer}

Let us now consider a thought experiment where a physical object falls towards the singularity. Our goal is to estimate the amount of energy the object encounters as it approaches $r=0$. To begin, we write the energy-momentum tensor for this metric as measured by a static observer (with respect to the matter source)
\begin{equation}\label{eq: energy-momentum tensor}
    T_{\mu \nu}=\begin{pmatrix}
\frac{m'(r)}{4\pi r^{2}}\left(1-\frac{2m(r)}{r}\right) & 0 & 0 & 0\\
0 & -\frac{m'(r)}{4\pi r^{2}} \frac{1}{\left(1-\frac{2m(r)}{r}\right)} & 0 & 0\\
0 & 0 & -\frac{m''(r)}{8\pi r} r^{2} & 0\\
0 & 0 & 0 & -\frac{m''(r)}{8\pi r} r^{2}\sin^{2}(\theta) 
\end{pmatrix}\,,
\end{equation}
To compute the energy density measured by a freely falling object, we evaluate $\rho=T_{\mu \nu}u^{\mu}u^{\nu}$, where $u$  is the four-velocity of the infalling object. We consider a radially infalling object whose trajectory is described by~\eqref{eq: four-velocity, radially infalling}. Using this, we now compute the energy density measured by the infalling object:
\begin{equation}\label{eq:density}
    \rho_{\rm free}=T_{\mu \nu}u^{\mu}u^{\nu}=\frac{m'(r)}{4\pi r^{2}}\,,
\end{equation}
which blows up for $m(r)\simeq r$ near $r=0$. However, while the local energy density increases unbounded, the total energy encountered by an infalling object over a finite volume remains finite.

To be concrete, let us consider a spherical object of radius $a=10$ meters, with the same density as water. The mass of this object is given by $m_{w}=\rho_{w}\frac{4}{3}\pi a^{3}\simeq 4.2\times10^{6}$ kg . Now, consider a sphere of the same size centered around the singularity. Using~\eqref{eq:density}, the mass enclosed within this region, as seen by the infalling object, is $m(a)$. With the mass function introduced before~\eqref{Eq:Integrable}, for a stellar-mass black hole with $M\sim10^{30}$ kg, we find that \mbox{$m(a=10m)\simeq 8.6*10^{27}$ kg}. So, this result suggests that, in a physical scenario, an extended object falling towards an integrable singularity will encounter an immense amount of energy (compared to its own one) as it approaches $r=0$. 

While the total energy remains finite, this does not necessarily imply that the object can pass through or even get arbitrarily close to the singularity without significant consequences. The sheer magnitude of the energy --- in this case almost $10^{21}$ times larger than the object's own mass --- raises serious concerns about the viability of traversing such a region. Even though tidal forces remain finite, the extreme energy encountered could lead to severe physical effects, such as rapid heating, structural disintegration, etc. 

This suggests that, despite its mathematical integrability, the singularity may still represent an effective physical barrier, challenging the assumption that an infalling object could safely reach or cross it. However, we should point out that the calculations presented here have been performed for a specific mass function, \eqref{Eq:Integrable}. In principle, one can adopt a different mass function, provided it satisfies the integrability condition \eqref{eq:integrabilitycondition}. Nevertheless, even in such cases, the energy encountered by an infalling observer will remain proportional to the parameters introduced in the chosen solution. This implies that one can always select a sufficiently low-density object that would still experience large energy transfer compared to its rest mass. In conclusion, while it is true that the encountered energy can be made relatively small for specific choices of parameters and for certain objects, it will always be large for a sufficiently low-density observer.

\section{Conclusions}
\label{Sec: Conclusions}

In this work, we set out to explore the physical viability of so-called integrable singularities as potential alternatives to standard black-hole interiors. Our investigation focused on two key aspects: first, the theoretical assumption that physical quantities associated with the source --- such as the energy density --- remain integrable at $r=0$; and second, the practical question of whether extended objects can traverse the singularity under realistic physical conditions. By examining how matter and test-field perturbations behave within this framework, we aimed to discern whether integrable singularities offer a genuinely milder route to resolving singularities or whether they succumb to non-integrability and fine-tuning issues in more realistic scenarios.

A key result of this work is our analysis of linear test-field perturbations near the singularity, inspired by prior studies on Schwarzschild interiors [e.g., \cite{doroshkevich1978spacetime}]. By examining spin-zero fields in a static, spherically symmetric integrable model, we found that although the background solution is crafted to ensure the integrability of certain source-derived quantities, this condition fails for generic perturbations. In particular, scalar modes either behave like $\ln{(r)}$ near $r=0$ or blow up with an inverse power of $r$. Crucially, when $l>0$, the energy density of these modes becomes non-integrable at $r=0$. This demonstrates that the theoretical ``integrability" built into the metric for source terms does not, in fact, guarantee integrability for field perturbations on the same geometry.

Another issue we pointed out is that parts of an extended object do not follow geodesics precisely. Instead, their motion is governed by the Mathisson-Papapetrou-Dixon (MPD) equations, which account for spin-curvature coupling and internal stresses. As a result, different points within an extended body can deviate from the finely tuned radial geodesic required for a weak singularity, with parts of the object inevitably moving along non-radial geodesics where the singularity is strong and destructive. This makes traversing an integrable singularity far less plausible than crossing a Cauchy horizon, where the singularity is weak for all geodesics.

To further probe the physical implications of these singularities, we considered a thought experiment in which a finite-sized object falls toward $r=0$. While tidal forces remain finite, the object experiences an enormous accumulation of energy as it approaches the singularity. This suggests that even in cases where geodesic deviation does not lead to immediate destruction, the energy flux encountered by a physical observer could still make the singularity effectively impassable, way before the object gets close to it.

Taken together, these results highlight several challenges associated with integrable singularities. While they offer a ``low-regularity'' extension of spacetime, their extreme fine-tuning, sensitivity to deviations from idealized motion, and the generic non-integrability of field perturbations, raise important questions about their physical viability. These issues suggest that integrable singularities require a more careful examination, particularly in the context of realistic astrophysical scenarios. Further research is needed to determine whether such singularities can remain well-behaved under more general conditions, or whether new mechanisms --- potentially involving quantum gravitational effects --- are necessary to ensure their stability.

\appendix
\section{Analysis of the solutions to the Frobenius series}\label{sec: appendix}

In Sec.~\ref{Subsec: perturbations}, we found the solutions to the indicial equation,~\eqref{eq: Spin zero, R}. As we mentioned, for $l=0$, the solutions simplify to $R_{1}=0$ and $R_{2}=1$. However, for $l>0$, we have
\begin{equation}\label{eq:roots, appendix}
    \begin{split}
        R_1=\frac{1}{2}\left(1-\underbrace{\frac{\sqrt{-4l(l+1)+\alpha}}{\sqrt{\alpha}}}_{C} \right),\quad 
        R_2=\frac{1}{2}\left(1+\underbrace{\frac{\sqrt{-4l(l+1)+\alpha}}{\sqrt{\alpha}}}_{C} \right)\,.
    \end{split}
\end{equation}

We are faced with two cases. If $2l\leq-1+\sqrt{1+\alpha}$, then $C$ is a positive real number, therefore $R_{1}\leq\frac{1}{2}$ and $R_{2}\geq\frac{1}{2}$, meaning that $R_{1}-1\leq -\frac{1}{2}$ and $R_{2}-1\geq -\frac{1}{2}$, which are relevant powers in the Frobenius series of the field~\eqref{eq: Spin zero, solution leading term}.

On the other hand, for $2l>-1+\sqrt{1+\alpha}$, $C$ is a purely imaginary number $C=i\ C'$, where $C'$ is a positive real number. Therefore, $R_{1}-1=-\frac{1}{2}-iC'$ and $R_{2}-1=-\frac{1}{2}+iC'$. This means that the two independent solutions to~\eqref{eq: Spin zero, solution} near $r=0$ would be
\begin{equation}\label{eq: Spin zero, roots, appendix}
    \begin{split}
        &\Psi_{0}^{1}=r^{\frac{1}{2}}\ \sin\left(C'\ln{(r)}\right)\sum_{n=0}^{\infty} a_{n}r^n\,,\\
        &\Psi_{0}^{2}=r^{\frac{1}{2}}\ \cos\left(C'\ln{(r)}\right)\sum_{n=0}^{\infty} a_{n}r^n\,,
    \end{split}
\end{equation}
which like before, for $n=0$
\begin{equation}\label{eq: Spin zero, field, appendix}
    \begin{split}
        &\Phi_{0}^{1}\propto r^{-\frac{1}{2}}\ \sin\left(C'\ln{(r)}\right)\,,\\
        &\Phi_{0}^{2}\propto r^{-\frac{1}{2}}\ \cos\left(C'\ln{(r)}\right)\,,
    \end{split}
\end{equation}
and since $-1<\sin{x}<1$ and $-1<\cos{x}<1$, therefore the field blows up as $\Phi_{0}\propto r^{-\frac{1}{2}}$ as $r\to0$. 

One can also perform the calculation of energy-density as in~\eqref{eq: Spin zero, energy-density}, keeping $\sin$ and $\cos$ functions, knowing that $\Phi_{0}\propto \Phi_{0}^{1}+\Phi_{0}^{2}$
\begin{equation}\label{eq: energy-density appendix}
        \rho\propto (\partial_{r} \Phi_{0})^2\propto\frac{1}{4r^{3}}\left[(2C'-1)\cos{\left(C'\ln{(r)}\right)}-(2C'+1)\sin{\left(C'\ln{(r)}\right)}\right]^2\,,
\end{equation}
which is indeed non-integrable
\begin{equation}\label{eq:non-integrable appendix}
    \begin{split}
        \int_{0}^{r}\rho\ r^{2}\sin{\theta}\, {\rm d}r\, {\rm d}\theta\, {\rm d}\phi&=\pi\left[\left(4C'^{2}+1\right)\ln{(r)}\ +\left(2C'-\frac{1}{2C'}\right)\cos{\left(2C'\ln{(r)}\right)\ -2\sin{\left(2C'\ln{(r)}\right)}}\right]\\ &=\infty\,.
    \end{split}
\end{equation}
confirming the fact that for $l>0$, the energy-density of the perturbations are always non-integrable.

\section*{Acknowledgments}

The authors wish to thank Roberto Casadio for useful remarks on an early version of the manuscript.
\bibliography{references}
\end{document}